\documentclass[conference]{IEEEtran}
\usepackage[ruled,vlined]{algorithm2e}

\ifCLASSINFOpdf

\else

\fi

\usepackage{amsfonts,epsfig,cite,array,multirow,graphicx,amsmath,ltablex,tabularx,setspace,amssymb,multirow}
\usepackage{schemabloc,amsthm}
\usepackage{subfig}
\usepackage[numbers,sort&compress]{natbib}
\usetikzlibrary{circuits, arrows}
\usepackage{verbatim}
\usepackage{graphicx}
\usepackage{epstopdf}
\usepackage{mathtools}
\usepackage{xcolor}
\usepackage{marvosym}
\usepackage{calc}

\usepackage{lipsum}
\usepackage{dblfloatfix}
\usepackage{color}
\usepackage{etoolbox}
\usepackage{epsfig}
\usepackage[inline]{enumitem}
\theoremstyle{plain}
\newtheorem{theorem}{Theorem}

\newtheorem{rem}{Remark}
\newtheorem{exm}{Example}

\usepackage{tikz}
\usepackage{collectbox}
\DeclareFontFamily{U}{mathx}{\hyphenchar\font45}
\DeclareFontShape{U}{mathx}{m}{n}{
	<5> <6> <7> <8> <9> <10>
	<10.95> <12> <14.4> <17.28> <20.74> <24.88>
	mathx10
}{}
\DeclareSymbolFont{mathx}{U}{mathx}{m}{n}
\DeclareMathSymbol{\bigtimes}{1}{mathx}{"91}
\makeatletter

\begin{document}
\tikzset{new spy style/.style={spy scope={%
			magnification=2.8,
			size=1.25cm,
			connect spies,
			every spy on node/.style={
				rectangle,
				draw,
			},
			every spy in node/.style={
				draw,
				rectangle,
			}
		}
	}
}
\SetKwRepeat{Do}{do}{while}
\newcolumntype{L}[1]{>{\raggedright\let\newline\\\arraybackslash\hspace{0pt}}m{#1}}
\newcolumntype{C}[1]{>{\centering\let\newline\\\arraybackslash\hspace{0pt}}m{#1}}
\newcolumntype{R}[1]{>{\raggedleft\let\newline\\\arraybackslash\hspace{0pt}}m{#1}}
\pgfdeclarelayer{background}
\pgfdeclarelayer{foreground}
\pgfsetlayers{background,main,foreground}
\newcommand{\oeq}{\mathrel{\text{\sqbox{$=$}}}}
\setlength{\textfloatsep}{0.1cm}
\setlength{\floatsep}{0.1cm}
\tikzstyle{int}=[draw, fill=white!20, minimum size=2em]
\tikzstyle{init} = [pin edge={to-,thin,black}]
\title{
\huge{OTFS-NOMA based on SCMA}
\vspace{-.3em} }

\author{
    \IEEEauthorblockN{\small{Kuntal Deka$^{1}$, Anna Thomas$^{1}$, and Sanjeev Sharma$^{2}$ }}\\
	$^1$Indian Institute of Technology Goa, India,   $^2$Indian Institute of Technology (BHU) Varanasi, India}

\maketitle

\begin{abstract}
Orthogonal Time Frequency Space (OTFS) is a $\text{2-D}$ modulation technique that has the potential to  overcome the challenges faced by  orthogonal frequency division multiplexing (OFDM) in high Doppler environments. 
 The  performance of OTFS  in a multi-user scenario with  orthogonal multiple access (OMA) techniques has been impressive.  Due to the requirement of massive connectivity in 5G and beyond, it is immensely essential to devise and examine the OTFS system  with the  existing  Non-orthogonal Multiple Access (NOMA)  techniques. 
 In this paper, we propose a multi-user  OTFS system  based on  a code-domain NOMA technique called Sparse Code Multiple Access (SCMA).  This system is referred to as the OTFS-SCMA model. The framework for OTFS-SCMA is designed for both downlink and uplink.  First, the sparse SCMA codewords are strategically placed on the delay-Doppler plane such that the overall overloading factor of the OTFS-SCMA system is equal to that of the underlying basic SCMA system. The receiver in downlink performs the detection in two sequential phases:  first, the conventional  OTFS detection using the method of  linear minimum mean square error (LMMSE), and then the conventional  SCMA detection. For uplink, we propose a single-phase detector based on  message-passing algorithm (MPA) to detect the multiple users' symbols. The performance  of the proposed OTFS-SCMA system is validated through extensive simulations both in downlink and uplink. We consider delay-Doppler planes of different parameters and  various SCMA systems of overloading factor up to 200$\%$. The performance of OTFS-SCMA is compared with those of existing OTFS-OMA techniques. The comprehensive investigation demonstrates the usefulness of   OTFS-SCMA in  future wireless communication standards. 
 
\end{abstract}


\IEEEpeerreviewmaketitle

 \vspace{-0.1in}
\section{Introduction}
Orthogonal Time Frequency Space (OTFS) has emerged as a promising modulation technique that  eliminates  the shortcomings of orthogonal frequency division multiplexing (OFDM) in 5G \cite{Hadani,Hadani2017a}.
 Many applications facilitated by 5G involve vehicle-to-vehicle and  vehicle-to-infrastructure communication scenario with the presence of high-speed vehicles.  OFDM  is not robust to  such  high Doppler environments   as the  sub-carriers don't remain  orthogonal.
 
  OTFS modulation technique succeeds in such cases by operating in the delay-Doppler domain instead of the conventional time-frequency domain.    By representing the  channel in delay-Doppler domain, it is possible to convert the time-variant channel response $h(\tau,t)$ to time-invariant channel response $h(\tau,\nu)$ where, $t, \tau$ and $\nu$ represent time, delay and Doppler respectively. This representation  directly describes the geometrical features of a channel. As the reflecting objects move slowly, the time-variation of the delay-Doppler channel representation is minimal.   Apart from the channel representation, the most distinguishing  feature of the OTFS modulation is that the information symbols are directly put in the delay-Doppler grid. 
 The OTFS modulation principle is inspired by  the existence of 2-D orthogonal signals in delay-Doppler domain  where they behave as  simultaneously localized in both time and frequency. An information symbol placed on a delay-Doppler grid  spans the entire allotted time-frequency plane using these orthogonal basis functions. Thus irrespective of the position  of symbol in the delay-Doppler grid, all of the symbols with the same power will experience the same channel. This invariance property provides OTFS its remarkable performance compared to existing modulation techniques especially in high Doppler channel conditions \cite{Hadani_2018_arxiv, Hadani}. 

Non-orthogonal multiple access (NOMA) is a popular multiple access framework particularly suited for applications involving massive connectivity \cite{dai2015non}.  The NOMA schemes can be divided into two domains:  \textit{power domain}  and \textit{code-domain}.  In power-domain NOMA, different users are identified using different levels of the assigned power \cite{islam2017power}. In code-domain NOMA, the multiple users are distinguished using different codewords assigned to them \cite{dai_survey_noma}.  Sparse Code Multiple Access (SCMA) is  a  code-domain NOMA technique where  sparse codewords are assigned to the users' symbols \cite{nikopour2013}.  The sparsity of the codewords facilitates the successful use of  the  message passing algorithm (MPA) \cite{sum_product} for the detection of the users' symbols. 


{\underline{\textit{Related work}}}:
The study of  multiple access technologies in the OTFS framework is an important topic.  
 The authors in \cite{Khammammetti2019a} proposed an OTFS-OMA scheme such that the multiple users' symbols are spread at equal interval over the entire delay-Doppler plane. These symbols are restricted to  non-overlapping contiguous sub-blocks in the time-frequency plane.  An OTFS-OMA scheme is presented in \cite{surabhi_MA} where the multiple users' symbols are put in non-overlapping sub-blocks of the delay-Doppler plane.
 The OTFS-OMA method in \cite{Augustine2019} allocates interleaved symbols of  different users in time-frequency domain. Note that here unlike \cite{Khammammetti2019a},  the allocation is not contiguous in the time-frequency plane, it is rather interleaved, i.e., between two symbols  of one user, a symbol from another user can be put. 
 The authors in \cite{Ding2019} considered a power-domain NOMA system  where the user
 with the highest velocity operates with OTFS and  the remaining users operate with OFDM.  In \cite{chatterjee2020non},  power domain NOMA is considered with OTFS. Link level simulations of the coded OTFS-power domain NOMA system are presented for downlink and uplink.

{\underline{\textit{Contributions}}}: In this paper, we propose an  OTFS-NOMA scheme based on a code-domain NOMA technique.  The code-domain NOMA considered here  is SCMA.  Although OTFS-NOMA schemes have been studied in power-domain, to the best of the authors' knowledge,  this work is the first attempt to build an OTFS-NOMA system  in the code-domain. First, we configure the OTFS-SCMA scheme in downlink.   The SCMA codewords are sparse column vectors. We  consider two schemes of allocating the SCMA codewords over the delay-Doppler plane.  First scheme allocates the vector codewords in the vertical sub-columns in the Doppler  direction. In the second scheme, the codewords are placed horizontally in the direction of delay bins.  The detector in downlink is a two-phased process: (1) OTFS detector followed by  (2) SCMA detector.   The BER performances of this OTFS-SCMA method  are evaluated for both the schemes with different sizes of delay-Doppler plane.  It is observed that the codeword allocation scheme is not crucial.  The BER performances are compared with those of existing conventional OTFS-OMA methods. In most of the cases, the OTFS-SCMA approach is found to provide better results than the existing OTFS-OMA.  Next we devise the strategy of OTFS-SCMA in uplink. It turns out that the simple two-stage detection approach adopted in downlink is not applicable   in uplink.  A combined approach is considered to carry out OTFS and SCMA detection. An  MPA detector is formulated  to detect the multiple users' data at one go.  The simulation results for different delay-Doppler planes are provided. 
 
 {\underline{\textit{Outline}}}:   Section~\ref{sec::prelim} describes the  preliminaries such as OTFS modulation principle and input-output relations, SCMA system model and it's related parameters. The proposed method of OTFS-SCMA is presented in Section~\ref{sec::prop}. The simulation results are presented and analyzed in Section~\ref{sec::simulations}.  Section~\ref{sec::conc} concludes the paper. 

{\underline{\textit{Notations}}}:  $\mathbb{C}$ denotes  the set of all complex numbers. $\mathbb{A}$  refers to  the underlying alphabet for the symbols. 
Boldface uppers-case  and boldface lower-case letters are used to denote the matrices and vectors respectively. 
$I_N$
denotes the identity matrix of size $N \times N$.   For a vector $\bf{h}$, $\text{diag} (\bf{h})$ is  the diagonal matrix with the first diagonal being  $\bf{h}$.   The complex conjugation of $x$ is denoted by $x^{\ast}$. For any integers $k$ and $N$, the notation $[k]_N$ refers to $k \mod{N}$.

\section{Preliminaries} \label{sec::prelim}
\subsection{OTFS} \label{OTFS}
This section briefly explains the  delay-Doppler symbol representation and the modulation-demodulation  in OTFS.
OTFS modulation technique introduced the idea of the allocation  of symbols in delay-Doppler domain.  In OTFS, the information symbols (e.g. QAM symbols) are arranged on a 2-D grid called delay-Doppler grid represented by $\Gamma_{M,N}$, where $M$ and $N$ correspond to the delay and Doppler dimension respectively. A total  of $MN$ QAM symbols can be transmitted on $\Gamma_{M,N}$.  The input signal  $x[k,l]$ refers to  the symbol at  $k^{\text{th}}$ Doppler bin and $l^{\text{th}}$ delay bin, where $k=0,1,...N-1$ and $l=0,1,...M-1$.
\subsubsection{Design parameters of delay-Doppler grid} Consider a data transmission frame of duration $T_{f}=NT$ and bandwidth $B=M\Delta f$ such that $T=\frac{1}{\Delta f}$. The delay-Doppler grid has to be designed in such a way that  $\tau_{\max}<\frac{1}{\Delta f}$ and $\nu_{\max}<\frac{1}{T}$ where $\tau_{\max}$ and $\nu_{\max}$ are the delay and the Doppler spread respectively. Therefore $\Gamma_{M,N}$ has a delay interval of $\Delta\tau=\frac{1}{M\Delta f}$ and Doppler interval of $\Delta\nu=\frac{1}{NT}$. Hence to transmit $NM$ symbols over a frame of duration $T_{f}$ and bandwidth $B$, the choice of $N$ and $M$ depends on the delay and Doppler conditions of the channel. A channel with a high Doppler spread $\nu_{\max}$ would require a  higher $N\Delta\nu$. It points to smaller $T$ and larger $\Delta f$, which directly implies larger $N$ and smaller $M$. Similarly if the channel has higher delay spread  $\tau_{\max}$, we require a higher $M\Delta\tau$ and the design of $\Gamma_{M,N}$ is to be done with a larger $M$ and a smaller $N$ \cite{ravi_2018_TWC}.

\begin{figure*}[t] 	
	\centering
	\includegraphics[height=3.7cm, width=17cm]{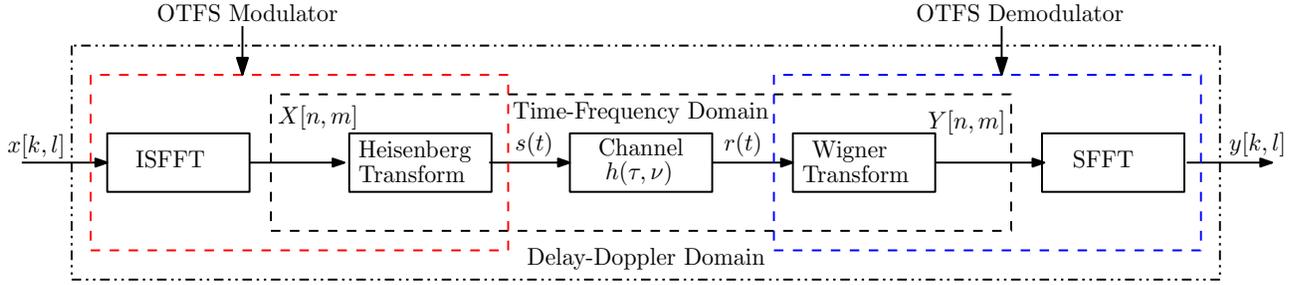}

	\caption{OTFS Block Diagram.}
	\label{otfs_block}
\end{figure*}
The basic modulation-demodulation involved in an OTFS scheme is shown in  Fig.~\ref{otfs_block}. It includes two stages of modulation/demodulation, one in the delay-Doppler domain and the other  in the time-frequency domain.
\subsubsection{Operations in Delay-Doppler Domain } 
 As we know, the QAM symbols $\left(x[k,l]\text{s}\right)$ to be transmitted are arranged in $\Gamma_{M,N}$ grid initially. The symbols $\left(y[k,l]\text{s}\right)$ finally received are also in $\Gamma_{M,N}$ grid. A 2-D transform pair known as Symplectic Finite Fourier Transform (SFFT) and Inverse Symplectic Finite Fourier Transform (ISFFT) is used in OTFS. SFFT is used for forward transformation from the  time-frequency  domain to the delay-Doppler domain and   ISFFT is used  for the reverse.  It is noteworthy that  SFFT  (ISFFT too) is a    combination of a forward Fourier transfom and an inverse Fourier transform \cite{Hadani_2018_arxiv,ravi_2018_TWC}.  At first the symbols in delay-Doppler domain are  transformed into  time-frequency domain using ISFFT as shown below: 
\begin{equation}
\label{eq:dis_isfft}
X[n,m] =  \frac{1}{NM}\sum_{k=0}^{N-1}\sum_{l=0}^{M-1}x[k,l]e^{j2\pi(\frac{nk}{N} - \frac{ml}{M})}.
\end{equation}
Similarly at the receiver side, the signals received in the time-frequency  domain are  finally transformed back to the  delay-Doppler domain using SFFT as described below:
\begin{equation}
\label{eq:dis_sfft}
y[k,l] =  \sum_{n=0}^{N-1}\sum_{m=0}^{M-1}Y[n,m]e^{-j2\pi(\frac{nk}{N} - \frac{ml}{M})}.
\end{equation}
Since the 2-D grid has to be limited to a dimension of $M\times N$, a rectangular windowing function is used both at the transmitter and the receiver side.
\subsubsection{Operations in Time-Frequency Domain } \label{sec::TF_des} The symbols $X[n,m]$s look like inputs to  regular OFDM data transmission. The only difference is that OTFS is a 2-D modulation technique while OFDM is 1-D modulation. An OTFS signal can be considered as a sequence of OFDM signals. The 2-D modulated  signal in the time-frequency domain has to be converted to time-domain using a suitable transformation.   Heisenberg transform deals with this transformation at the transmitter side using a basis pulse represented by $g_{\text{tx}}(t)$.  Heisenberg's transform can be mathematically expressed as the  modulation of $X[n,m]$ on $g_{\text{tx}}(t)$ as shown below:
\begin{equation}
\label{eq:tx_pulse}
s(t) = \sum_{n=0}^{N-1}\sum_{m=0}^{M-1}X[n,m]e^{j2\pi m\Delta f(t-nT)}g_{\text{tx}}(t-nT).
\end{equation}
The signal $s(t)$ is transmitted over the wireless channel with impulse response $h(\tau,\nu)$.  The received signal $r(t)$ is given by 
\begin{equation}
\label{eq:tx_rx}
r(t) = \iint h(\tau,\nu)e^{j2\pi\nu(t-\tau)}s(t-\tau)\,d\tau d\nu  +n(t)
\end{equation}
where, $n(t)$  is the additive white Gaussian noise (AWGN) signal. 

Similar to Heisenberg transform, there  exists an inverse transform at the receiver side for transforming  the  time-domain signal back to the time-frequency plane. This inverse transform is known as Wigner transform which demodulates the received signal $r(t)$ to $Y[n,m]$. 
A suitably designed  basis pulse $g_{\text{rx}}(t)$ is used in the received side. 

For demodulation, the matched filter first computes the cross-ambiguity function  $A_{g_{\text{rx}},r}(\tau,\nu)$  between the  received signal $r(t)$ and the receive pulse $g_{\text{rx}}(t)$.  
The cross-ambiguity function is a type of 2-D correlation function and is  given by
\begin{equation}
\label{eq:cross_amb}
A_{g_{rx},r}(\tau,\nu) \triangleq \int e^{-j2\pi\nu(t-\tau)}g_{\text{rx}}^{\ast}(t-\tau)r(t) \, dt.
\end{equation}
Sampling the matched filter output at regular intervals of $(nT,m\Delta f)$ gives the 2-D time-frequency  demodulated signal $Y[n,m]$ as given below:
\begin{equation}
\label{eq:match_filter}
Y[n,m] = A_{g_{rx},r}(\tau,\nu) \arrowvert_{\tau=nT,\nu=m\Delta f}.
\end{equation}

The selection of $g_{\text{tx}}(t)$ and $g_{\text{rx}}(t)$  is to be done wisely so that ideally they obey the  bi-orthogonality property otherwise there will be interference from different delay and Doppler bins.
The bi-orthogonality property can be written as given in the following:
\begin{equation}
\label{eq:bi_ortho}
\int e^{-j2\pi m\Delta f(t-nT)}g_{\text{rx}}^{*}(t-nT)g_{\text{tx}}(t) \, dt = \delta(m)\delta(n).
\end{equation}
The pulses $g_{\text{tx}}(t)$ and $g_{\text{rx}}(t)$ which satisfy the bi-orthogonality property are known as \textit{ideal} pulses. For the simple \textit{rectangular}  pulses, $g_{\text{tx}}(t)$ and $g_{\text{rx}}(t)$ have amplitude $1/{\sqrt{T}}$ for $t \in [0, T]$  and 0 at all other values. Note that the rectangular pulses don't satisfy the bi-orthogonality condition.  The ideal pulses don't exist in real world. Therefore we focus on the rectangular pulses in the rest of the paper.  
\subsubsection{Input-Output Relation}
The input-output relation refers to the relation between $x[k,l]$s and $y[k,l]$s. The input-output relation is helpful for designing detectors in the receiver side. 
The channel response in the delay-Doppler domain can be sparsely represented as
\begin{equation}
\label{eq:h_sparse}
h(\tau,\nu) = \sum_{i=1}^{P}h_{i}\delta(\tau - \tau_{i})\delta(\nu - \nu_{i})
\end{equation}
where,  $P$ denotes the number of paths in the channel; $h_i$, $\tau_i$ and $\nu_i$ denote the $i^{\text{th}}$ path's  gain, delay and Doppler shift respectively.  To conveniently express the input-output relation, the delay and Doppler taps for the $i^{\text{th}}$ path can be alternatively represented as \cite{ravi_2018_TWC}
\begin{equation*}
\tau_i=\frac{l_{\tau_i}}{M\triangle f},\;\; \nu_i=\frac{k_{\nu_i}+\kappa_{\nu_i}}{NT}
\end{equation*}
with some integers $l_{\tau_i},k_{\nu_i}$ and   $\kappa_{\nu_i} \in \left[-0.5,0.5\right]$ which represents fractional Doppler shift.

With the above sparse channel representation, the received signal $y[k,l]$ in the delay-Doppler domain for rectangular pulses is given by \cite{ravi_2018_TWC}
\begin{equation}
\label{eq:otfs_io}
\begin{aligned}
y[k,l]\approx \sum_{i=1}^{P} \sum_{q=-N_i}^{N_i} h_i& e^{j2\pi \left(\frac{l-l_{\tau_i}}{M}\right)\left(\frac{k_{\nu_i}+\kappa_{\nu_i}}{N}\right)} \alpha_i\left(k,l,q\right)\\
& \times x\left[\left[k-k_{\nu_i}+q\right]_N, \left[l-l_{\tau_i}\right]_M\right]
\end{aligned}
\end{equation}
where,
\begin{equation}
\label{eq:otfs_io_sub}
\begin{aligned}
\alpha_i(k,l,q) & =\begin{cases}
\frac{1}{N}\beta_i(q) &  l_{\tau_i}\leq l <M\\    
\frac{1}{N}\left(\beta_i(q) -1\right)e^{-j2\pi \frac{\left[k-k_{\nu_i}+q\right]_N}{N}} & 0 \leq l <l_{\tau_i}
\end{cases}\\
\beta_i(q)&=\frac{e^{-j2\pi \left(-q-\kappa_{\nu_i}\right)}-1}{e^{-j2\frac{\pi}{N} \left(-q-\kappa_{\nu_i}\right)}-1}.
\end{aligned}
\end{equation}
 In (\ref{eq:otfs_io})  and (\ref{eq:otfs_io_sub}), the number $N_i$  appears when   fractional Doppler shift exists.  $N_{i}$ denotes the number of neighboring Doppler points interfering with the particular point under consideration. Usually the value of  $N_i$ is  significantly smaller than the total number $N$ of Doppler bins.  Observe that the AWGN term is omitted in (\ref{eq:otfs_io}) for compact  description. 
 
 The input-output relation given in (\ref{eq:otfs_io}) can be represented by a set of linear equations of the following form \cite{ravi_2018_TWC}:
\begin{equation}
\label{eq:matrix_sparse}
\mathbf{y = Hx+z}
\end{equation}
where, $\mathbf{y}\in \mathbb{C}^{NM \times 1}$, $\mathbf{z} \in \mathbb{C}^{NM \times 1}$ and $\mathbf{x} \in \mathbb{A}^{NM \times 1}$ are the row-wise vectorized versions of the output, AWGN and the input respectively;  $\mathbf{H} \in \mathbb{C}^{NM \times NM}$ is called as the coefficient matrix. A detailed explanation on formulating $\mathbf{H}$ matrix with any practical waveforms is given in \cite{Raviteja2019a}. This linear relationship is valid in the case of both ideal and practical pulse, though the elements  of $\mathbf{H}$ matrix change. The coefficient  matrix  $\bf{H}$ has exactly  $S=\sum_{i=1}^P (2N_i+1)$   non-zero elements in each row and column. In absence of fractional Doppler shifts (i.e. $N_i=0$), we have $S=P$. 
 This makes $\mathbf{H}$ matrix highly sparse. Owing to the  sparsity  of $\mathbf{H}$, the detection can be carried out with the help of  MPA \cite{ravi_2018_TWC,RavitejaWCNC,surabhi_MA}.   In place of MPA, other low-complexity linear equalization algorithms like LMMSE can also be effectively used by considering the block-circulant property of $\mathbf{H}$ matrix \cite{Tiwari2019,Surabhi2020a}.     

\subsection{SCMA}\label{SCMA}
SCMA is a highly sophisticated code-domain NOMA technique. 
A distinct codebook is assigned to each user such that no two users would have to use the  same codeword. 
The basic parameters of SCMA and the other related concepts are  briefly explained next.
\subsubsection{SCMA Parameters} An SCMA system  is represented as $J\times K$ model, where there are $J$ users sharing $K$  orthogonal resources. As $J>K$, we have an overloaded system with an  overloading factor, $\lambda=\frac{J}{K}>100\%$. 
\begin{figure}[!ht] 	
	\centering
	\includegraphics[height=6cm, width=9cm]{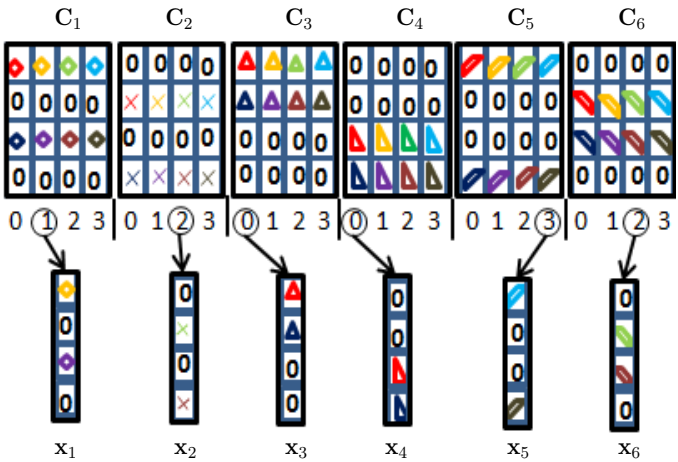}
	\caption{Basic $6\times 4$ SCMA Model.}
	\label{scma_block}
\end{figure}
As an example, the  $6\times 4$ SCMA system shown in Fig.~\ref{scma_block} has 6 users sharing 4 resources with an overloading factor of $150\%$.  The system has six distinct  codebooks $\{\mathbf{C_{1},C_{2},..., C_{6}}\}$ dedicated for each user. Each codebook consists of four different complex codewords representing four different information symbols $\{0,1,2,3\}$. Depending on the input data, one codeword  is selected by each user denoted by $\{{\bf{x}}_{1},{\bf{x}}_{2},..., {\bf{x}}_{6}\}$  for transmission where, ${\bf{x}}_j\in {\mathbb{C}}^{4\times 1 }$, $j=1,\ldots, 6$. The performance  of an SCMA system  is highly sensitive to the  codebooks. 
\subsubsection{Factor Graph} 
\begin{figure}[!ht] 	
	\centering
	\includegraphics[height=2.2cm, width=7cm]{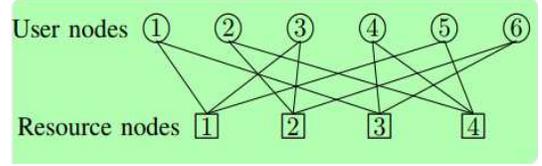}
	\caption{Factor graph for $J=6$, $K=4$ SCMA system.} 
	\label{fg}
\end{figure}

Observe from Fig.~\ref{scma_block} that the codewords of any particular codebook are sparse and contain 0s in specific locations. For any user, if a codeword has a non-zero element in the $j^{\text{th}}$ location, then it implies that the user is occupying the $j^{\text{th}}$ resource.    The sharing of resources amongst multiple users  can be  graphically represented by a factor graph. The factor graph for a $J\times K$ SCMA system contains $J$ user nodes and $K$ resource/factor nodes.  An edge is assigned between the $j^{\text{th}}$  user node  and the  $k^{\text{th}}$ resource node  only if the $j^{\text{th}}$ user occupies the   $k^{\text{th}}$ resource.    Fig.~\ref{fg} shows the factor graph for the SCMA model described in Fig.~\ref{scma_block}.  The degrees of a user node and a resource/factor node are denoted by  $d_{v}$ and $d_{f}$ respectively. For the factor graph shown in Fig.~\ref{fg}, we have $d_{v}=2$ and  $d_{f}=3$. 
An alternative of the  factor graph is the  factor matrix, $\mathbf{F}$. Each edge in the factor graph is denoted by a `1' in the factor matrix.  
For the factor graph shown in Fig.~\ref{fg}, $\mathbf{F}$ is obtained as given below:
\begin{equation}
\label{eq:F_matirx}
\mathbf{F=}
\begin{bmatrix}
1&0&1&0&1&0\\
0&1&1&0&0&1\\
1&0&0&1&0&1\\
0&1&0&1&1&0
\end{bmatrix}.
\end{equation}
The sparsity of factor graph or matrix  indicates that MPA can be used in SCMA detection. It is extensively reported in literature that the   MPA-based  detection provides excellent performance in SCMA \cite{nikopour2013,dai_2017}. 
 \subsubsection{Downlink}
 In downlink, the base station (BS) first sums up the codewords of all  $J$ users. This superimposed signal is broadcast to every user. 
 The received signal ${\mathbf{y}}_i=\left[y_{i1}, \ldots,  y_{iK}\right]^T$ at the $i^{\text{th}}$ user can be expressed as 
 \begin{equation}\label{downlink_SCMA}
 \mathbf{y}_i=\text{diag}\left({\bf{h}}_i\right)\sum_{j=1}^{J}   \mathbf{x}_{j}+\mathbf{n}_i
 \end{equation} 
 where,  ${\bf{h}}_i=\left[h_{i1}, h_{i2}, \ldots, h_{iK}\right]$ is the channel impulse response  vector between the BS and  the $i^{\text{th}}$ user, $\mathbf{n}_i$ is the   AWGN at the $i^{\text{th}}$ user and is complex  Gaussian distributed, i.e., $\mathbf{n}_i\sim \mathcal{CN}(0,N_{0}\mathbf{I}_K)$.
 \subsubsection{Uplink}
In uplink, each user incorporates  the respective channel to transmit the  information  to the BS. The BS receives the    signal $\textbf{y}$ which is given by 
\begin{equation}\label{uplink_SCMA}
\textbf{y}=\sum_{j=1}^{J}  \operatorname{diag(\textbf{h}_{j})}\textbf{x}_{j}+\textbf{n}
\end{equation}
where, $\mathbf{n}$ is the   AWGN at the BS. 
\section{Proposed Method of SCMA-based OTFS-NOMA}  \label{sec::prop}
In this section we propose the system model for OTFS-SCMA  for downlink and uplink.   The operations to be carried out  in the transmitter and the receiver side are described in details in the following. 
\subsection{In Downlink}
\begin{figure*}[t] 	
	\centering
	\includegraphics[height=4cm, width=17cm]{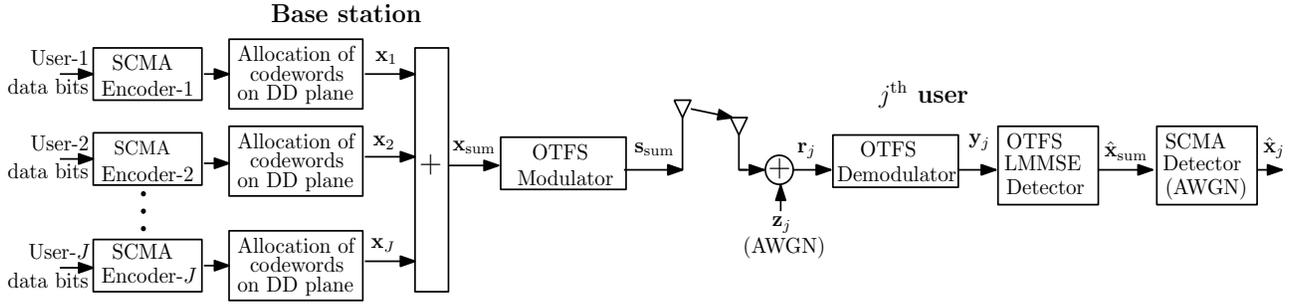}
	\caption{Block diagram of  OTFS-SCMA in downlink.}
	\label{block_downlink}
\end{figure*}
\begin{figure}[!ht] 	
	\centering
	\includegraphics[height=6.3cm, width=6cm]{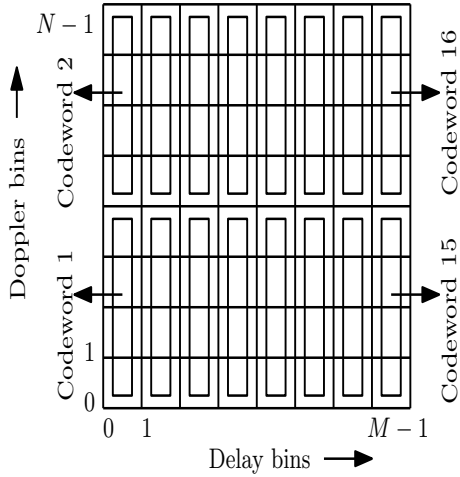}

	\caption{Allocation of SCMA codewords of length $K=4$ in delay-Doppler plane $\Gamma_{8,8}$, \textit{Scheme 1}.}
	\label{DD1}
\end{figure}

Consider a  downlink scenario with  $J$ receiving users and one transmitting BS as shown in Fig.~\ref{block_downlink}. Note that  only the $j^{\text{th}}$ user's receiver is depicted.  A $J\times K$   SCMA system is considered with overloading factor $\lambda=\frac{J}{K}$.   The length of a complex SCMA codeword is $K$. For allocation of the SCMA codewords over the delay-Doppler plane, we consider two schemes:

\textit{\underline{Scheme 1}}: 
 The SCMA codewords for the $j^{\text{th}}$ user, $j=1, \ldots ,  J$  are   allocated in the delay-Doppler plane $\Gamma_{M,N}$ along the Doppler direction in blocks of size $K\times 1$. This scheme is depicted in Fig.~\ref{DD1} for $M=8, N=8$ and $K=4$.   

\begin{figure}[!ht] 	
 	\centering
 	\includegraphics[height=6cm, width=6cm]{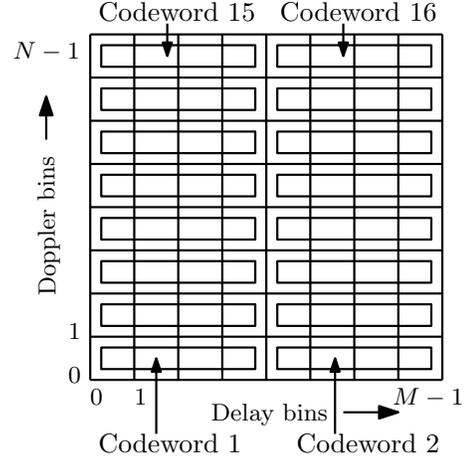}

 	\caption{Allocation of SCMA codewords of length $K=4$ in delay-Doppler plane $\Gamma_{8,8}$, \textit{Scheme 2}.}
 	\label{DD2}
 \end{figure}

 \textit{\underline{Scheme 2}}: 
 The SCMA codewords for the $j^{\text{th}}$ user, $j=1, \ldots ,  J$  are   allocated in the delay-Doppler plane $\Gamma_{M,N}$ along the delay direction in blocks of size $1\times K$. This scheme is depicted in Fig.~\ref{DD2} for $M=8, N=8$ and $K=4$.  
 
 The $J\times K$ SCMA system is referred to as the \textit{basic SCMA system} as it is repeated over the delay-Doppler grid multiple times.
 In the following, we present the method for Scheme~1.  The notations and the methods  can be easily extended to  Scheme 2.

\begin{figure}[!ht] 	
     	\centering
     	\includegraphics[height=2cm, width=9cm]{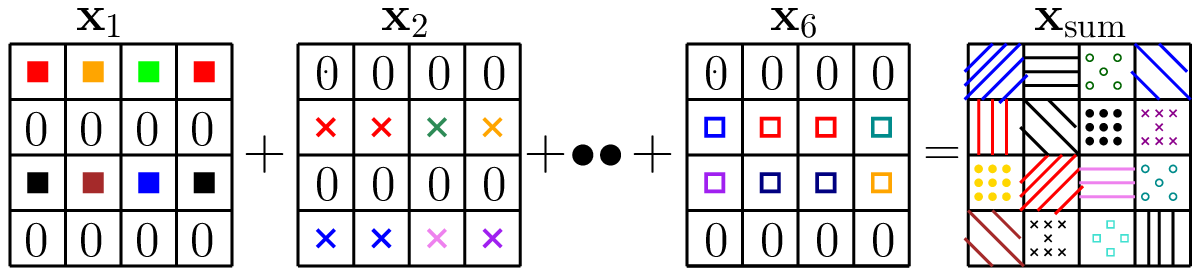}
     	\caption{Superposition  of SCMA codewords of all users in delay-Doppler plane $\Gamma_{4,4}$ with $6\times4$ SCMA system.}
     	\label{DD_downlink}
     \end{figure}

There are $M\times N$  slots for complex numbers  in   $\Gamma_{M,N}$. Thus, for every user, the total number of  symbols that can be accommodated inside $\Gamma_{M,N}$  is $N_{\text{sym}}=\frac{M\times N}{K}$ as the length of any SCMA codeword is $K$. 
Suppose ${\bf{x}}_j \in {\mathbb{C}}^{N\times M}$ is the input delay-Doppler frame for $j^{\text{th}}$ user.  The superimposed input signal  is given by
 \begin{equation}
 	{\bf{x}}_{\text{sum}}=\sum_{j=1}^{J}{\bf{x}}_j
 \end{equation}
  where, the summation is done bin-wise.  The superposition of the SCMA codewords of all users is depicted in Fig.~\ref{DD_downlink} for $\Gamma_{4,4}$ and $6\times 4$ SCMA system. Note that the delay-Doppler plane of each user contains a significant number of 0s as the SCMA  codewords  are sparse.  
The superimposed signal ${\bf{x}}_{\text{sum}}$
  goes through the OTFS modulator to produce the vector ${\bf{s}}_{\text{sum}} \in {\mathbb{C}}^{MN\times 1}$.
 The BS transmits ${\bf{s}}_{\text{sum}}$  and signal ${\bf{r}}_j$ received by the $j^{\text{th}}$ user is passed through an OTFS demodulator. Suppose the output of the OTFS demodulator is ${\bf{y}}_j$.   
 
 Invoking the input-output relationship between the BS and the $j^{\text{th}}$ user,  ${\bf{y}}_j$ can be written in terms of the superposition ${\bf{x}}_{\text{sum}}$ of the input SCMA codewords.   
  Suppose ${\bf{x}}_{\text{sum,vec}} \in {\mathbb{C}}^{MN\times 1}$ and ${\bf{y}}_{j,\text{vec}} \in {\mathbb{C}}^{MN\times 1}$ are the row-wise vectorized versions of ${\bf{x}}_{\text{sum}} \in {\mathbb{C}}^{N\times M}$ and  ${\bf{y}}_{j} \in {\mathbb{C}}^{N\times M}$ respectively. The effective input-output relation now becomes
  \begin{equation}
  \label{downlink_eq_OTFS_SCMA}
  {\bf{y}}_{j,\text{vec}}={\bf{H}}_j{\bf{x}}_{\text{sum,vec}}+{\bf{z}}_j
  \end{equation}
  where, ${\bf{H}}_j$ is the $NM\times NM$ complex coefficient matrix specifying the relationship between the input   ${\bf{x}}_{\text{sum,vec}}$ and output ${\bf{y}}_{j,\text{vec}}$,  and ${\bf{z}}_j \in {\mathbb{C}}^{MN\times 1}$  is the complex AWGN.

As the BS is transmitting $J\times N_{\text{sym}}$  symbols over $M\times N$  resources, the overloading factor of the OTFS-SCMA scheme becomes $\lambda=\frac{J\times N_{sym}}{M\times N}=\frac{J}{K}$. Thus the overloading factor of the OTFS-SCMA system is exactly equal to the underlying basic SCMA system.

    Observe from (\ref{downlink_eq_OTFS_SCMA}) that in order to recover ${\bf{x}}_{\text{sum,vec}}$ from ${\bf{y}}_{j,\text{vec}}$, any OTFS detector can be used. We propose to use a simple LMMSE-based detector  for the same. The estimate of ${\bf{x}}_{\text{sum,vec}}$    is obtained as 
    \begin{equation}
    \label{LMMSE}
    {\hat{\bf{x}}}_{\text{sum,vec}}={\bf{H}}_j^{\ast}\left[{\bf{H}}_j{\bf{H}}_j^{\ast}+\sigma_n^2{\bf{I}}_{MN}\right]^{-1}{\bf{y}}_{j,\text{vec}}.
    \end{equation}
    
    Note that (\ref{LMMSE})  involves matrix inversion which increases the computational complexity. In order to lower the complexity, the method proposed in \cite{Tiwari2019} may be used for OTFS detection.  The output ${\hat{\bf{x}}}_{\text{sum,vec}}$  of the LMMSE detector is fed to an MPA-based SCMA  detector block to obtain the estimate ${\hat{\bf{x}}}_j$ of the $j^{\text{th}}$ user's data ${\bf{x}}_j$.  Note that the SCMA-detection block contains $N_{\text{sym}}$ basic SCMA detectors.  Moreover, as the effects of fading have been canceled by the OTFS detector, we consider a simple AWGN channel for the SCMA detector. 
   
\subsection{In Uplink} \label{sec::uplink_prop}
\begin{figure*}[t] 	
	\centering
	\includegraphics[height=4cm, width=16cm]{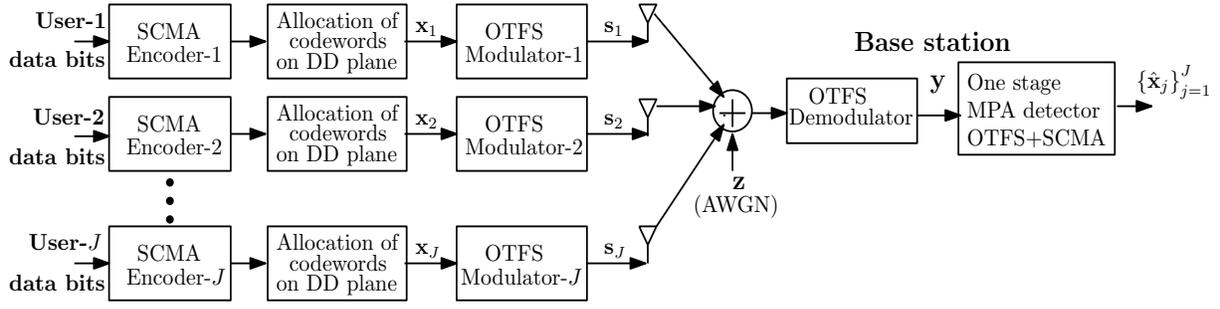}
	\caption{Block diagram of  OTFS-SCMA in uplink.}
	\label{block_uplink}
\end{figure*}

Consider an uplink scenario with $J$ transmitting users and one receiving BS as shown in Fig.~\ref{block_uplink}. The description of the methods pertains to Scheme 1. The SCMA codewords are allocated as  depicted in Fig.~\ref{DD1}. 
There are $J$  input-output relations between the output of the OTFS demodulator ${\bf{y}}$ and inputs ${\bf{x}}_j$ of each user's modulator, $j=1,\ldots,J$.  Combining these $J$  relations, the vectorized version  ${\bf{y}}_{\text{vec}}$ of   ${\bf{y}}$ can be expressed as                      
\begin{equation}
{\bf{y}}_{\text{vec}}=\sum_{j=1}^{J}{{\bf{H}}_j {\bf{x}}_{j,\text{vec}}}+ {\bf{z}}
\label{uplink_eq}
\end{equation}
where, ${\bf{z}} \in {\mathbb{C}}^{MN\times 1}$  is the complex AWGN.

Unlike in downlink, the receiver in uplink involves more than one $\bf{H}$ matrices . Therefore, the previously-discussed two-stage method of detection cannot be adopted here.  
From (\ref{uplink_eq}), ${\bf{y}}_{\text{vec}}$ can be written as 
\begin{equation}
\label{uplink_eq1}
{\bf{y}}_{\text{vec}}= {\bf{H}}_{\text{all}} {{\bf{x}}}_{\text{all}} + {\bf{z}}
\end{equation}
where, 
\begin{equation}
\label{all_eq}
\begin{aligned}
	{\bf{H}}_{\text{all}}& =\left[ {\bf{H}}_1, {\bf{H}}_2 ,\ldots, {\bf{H}}_J\right]   \text{ and}  \\
	{{\bf{x}}}_{\text{all}}& =\left[{\bf{x}}_{1,\text{vec}}, {\bf{x}}_{2,\text{vec}}, \ldots, {\bf{x}}_{J,\text{vec}}\right]^T
\end{aligned}
\end{equation}
Observe from (\ref{all_eq}) that ${\bf{H}}_{\text{all}}$ is an $MN\times JMN$ complex matrix and ${{\bf{x}}}_{\text{all}}$ is the information vector of length $JMN$.  However, as ${{\bf{x}}}_{\text{all}}$ contains  SCMA codewords, it will be a sparse vector.  The number of 0s in ${{\bf{x}}}_{\text{all}}$ is $\frac{JMNd_v}{K} $, where $d_v$ is the number of non-zero components in a length-$K$ codeword of an $J\times K$ SCMA system.  For an SCMA system with factor graph shown in   Fig.~\ref{fg}, ${{\bf{x}}}_{\text{all}}$ contains $3MN$ non-zero complex numbers.  Let ${{\bf{x}}}_{\text{all,compr}}$ denote  the compressed input vector after removing the 0s in ${{\bf{x}}}_{\text{all}}$. Similarly, let ${\bf{H}}_{\text{all,compr}}$
denote the effective compressed ${\bf{H}}$ matrix after deleting the columns corresponding to the locations of the 0s in ${{\bf{x}}}_{\text{all}}$.  Then, the input-output relationship in  (\ref{uplink_eq1})  can now be written as
\begin{equation}
\label{uplink_eqn_final}
{\bf{y}}_{\text{vec}}= {\bf{H}}_{\text{all,compr}} {{\bf{x}}}_{\text{all,compr}} + {\bf{z}}
\end{equation}
where, ${\bf{H}}_{\text{all,compr}} \in {\mathbb{C}}^{MN\times \frac{JMNd_v}{K}  } $  and $ {{\bf{x}}}_{\text{all,compr}} \in  {\mathbb{C}}^{ \frac{JMNd_v}{K} \times 1 }$. 

Observe from (\ref{uplink_eqn_final}), that the we have less observations than the number of variables. Therefore, a powerful detector must be employed in the receiver.  We consider  a single stage of MPA for the detection.  Note that the consecutive $d_v$ elements of $ {{\bf{x}}}_{\text{all,compr}} $  
are the non-zero elements of a particular SCMA codeword.  These $d_v$ elements should be considered together  as a single entity or  observation node as they correspond to a particular information symbol in the input side.   
 
 In (\ref{uplink_eqn_final}), we consider ${{\bf{x}}}_{\text{all,compr}}\in {\mathbb{C}}^{\frac{JMNd_v}{K}\times 1}$ as a collection of $\frac{JMN}{K}$ variable nodes. We call them variable nodes (not user nodes)  as they  refer to different symbols of the same user. Each variable node comprises of $d_v$ complex numbers but they all refer to one particular symbol. Similarly, in every row of ${\bf{H}}_{\text{all,compr}}$, the consecutive $d_v$ complex coefficients must be clubbed together.  With these new conventions, 
 (\ref{uplink_eqn_final}) can be written as

 \begin{equation}
 \small
 \label{uplink_eqn_final1}
 {\bf{y}}_{\text{vec}}= \begin{bmatrix}
  y_1\\  
  \vdots\\
  y_{MN}
  \end{bmatrix}
  = \begin{bmatrix} 
                {\bf{h}}_{1,1}    & \cdots & {\bf{h}}_{1,\frac{JMN}{K}}\\
                \vdots     & \cdots &  \vdots\\
                {\bf{h}}_{MN,1}   & \cdots & {\bf{h}}_{MN,\frac{JMN}{K}}
                \end{bmatrix}
  \begin{bmatrix} 
  {\bf{x}}_{1}\\  
  \vdots\\
  {\bf{x}}_{\frac{JMN}{K}}
  \end{bmatrix}
   + \begin{bmatrix} 
   {{z}}_{1}\\   
   \vdots\\
   {{z}}_{MN}
   \end{bmatrix}.
 \end{equation}
 
 In (\ref{uplink_eqn_final1}), for any $d$  and $c$, ${\bf{h}}_{dc} \in {\mathbb{C}}^{1\times d_v}$  and ${\bf{x}}_c \in {\mathbb{C}}^{d_v \times 1}$.  The corresponding factor graph  comprises of $MN$ observation nodes and $\frac{JMN}{K}$  variable  nodes.  An edge is assigned between the observation node $y_d$  and the variable  node ${\bf{x}}_c$  if ${\bf{h}}_{dc} \neq \left[0,\cdots, 0\right]_{1\times d_v}$.   The set of variable nodes connected to the $d^{\text{th}}$ observation node is denoted by ${\cal{M}}_d$   and  ${\cal{M}}_d^c={\cal{M}}_d\setminus \left\{c\right\}$. Similarly, ${\cal{N}}_c$  denotes the set of observation nodes connected to the $c^{\text{th}}$ variable node and ${\cal{N}}_c^d={\cal{N}}_c\setminus \left\{d\right\}$.  The MPA-based detection is carried out over this factor graph. Algorithm~\ref{algo:uplink}  presents the detailed procedures of the detection process. In MPA, the message update from an observation node is very critical and this step is further explained with the help one example furnished below.
 \begin{algorithm}[!htbp]
 	\caption{\small MPA-based detection  for OTFS-SCMA  in uplink }
 	\label{algo:uplink}
 	\small
 	\SetKwData{Left}{left}
 	\SetKwData{This}{this}
 	\SetKwData{Up}{up}
 	\SetKwFunction{Union}{Union}
 	\SetKwFunction{FindCompress}{FindCompress}
 	\SetKwInOut{Input}{input}
 	\SetKwInOut{Output}{output}
 	\Input{${\bf{y}}_{\text{vec}}$,   ${\bf{H}}_{\text{all,compr}}$, Alphabets $\left\{{\mathbb{A}}_j\right\}_{j=1}^J$,  $\frac{E_b}{N_0}$}
 	\Output{Estimated data symbol for  ${\hat{\bf{x}}}_{\text{all,compr}}$  }
 	\BlankLine 	
 	%
 	Initialization:  For every $c \in \left\{1,\ldots JMN/K\right\}$ and for every $d\in {\cal{N}}_c $, set  $V_{c\rightarrow d}^{(0)}\left({\bf{x}}_{jm}\right)=1/ A, m=1, \ldots, A   $, where $A$ is the size of the alphabet, Iteration index $l \leftarrow 0$\;
 	
 	\While{termination criteria not fulfilled}{
 		\For{$d\leftarrow 1$ \KwTo $MN$}{
 			 \ForEach{$c  \in  {\cal{M}}_d$}{
 			Compute $U_{d \rightarrow c}^{(l)}({\bf{x}}_{cm})$ for $m=1,\ldots A$
 					\begin{equation*}\begin{split}
 					& U_{d \rightarrow c}^{(l)}(m) = \sum_{{\bf{v}} \in {\cal{V}}_{dc} } \frac{1}{\pi N_0} \exp [-\frac{1}{N_0}| y_d -
 					{\bf{h}}_{dc}{\bf{x}}_{cm} \\ &
 					 - \sum_{c' \in {\cal{M}}_d^c} {\bf{h}}_{dc'}{\bf{v}}_{c'} |^2] \prod_{c' \in {\cal{M}}_d^c} V_{c'\rightarrow d}^{(l-1)}\left({\bf{v}}_{c'}\right)
 				\end{split}\end{equation*}
 				where,  $ {\cal{V}}_{dc}= \bigtimes\limits_{c' \in {\cal{M}}_d^c} {\mathbb{A}}_{c'}$ is the Cartesian product of alphabets with  ${\bf{v}} =\left[{\bf{v}}_{c'}\right]_{c' \in {\cal{M}}_d^c} $ is a member of ${\cal{V}}_{dc}$ and ${\bf{v}}_{c'} \in {\mathbb{A}}_{c'}$
 				$${\text{Normalization }} \;\; U_{d \rightarrow c}^{(l)}({\bf{x}}_{cm})=\frac{U_{d \rightarrow c}^{(l)}({\bf{x}}_{cm})}{\sum_{m'=1}^{A}U_{d \rightarrow c}^{(l)}({\bf{x}}_{cm'})}$$
 			}	}
 			
 			\For{$c\leftarrow 1$ \KwTo $JMN/K$}{
 				\ForEach{$d  \in  {\cal{N}}_c$}{
 					Compute $V_{c \rightarrow d}^{(l)}({\bf{x}}_{cm})$ for $m=1,\ldots A$
 					$$V_{c\rightarrow d}^{(l)}\left({\bf{x}}_{cm}\right) =\prod_{d' \in {\cal{N}}_c^d} U_{d' \rightarrow c}^{(l)}({\bf{x}}_{cm})$$ 
 					$${\text{Normalization }} \;\; V_{c \rightarrow d}^{(l)}({\bf{x}}_{cm})=\frac{V_{c \rightarrow d}^{(l)}({\bf{x}}_{cm})}{\sum_{m'=1}^{A}V_{c \rightarrow d}^{(l)}({\bf{x}}_{cm'})}$$
 				}
 			}
 			\For{$c\leftarrow 1$ \KwTo $JMN/K$}{
 				Compute $V_{c }^{(l)}({\bf{x}}_{cm})$ for $m=1,\ldots, A$
 					\begin{equation*}
 					\label{posterior_update}
 					V_{c}\left({\bf{x}}_{cm}\right)=\prod_{k \in {\cal{N}}_c} U_{d \rightarrow c}^{(l)}({\bf{x}}_{cm}).
 					\end{equation*}

 			}
 			If the posterior probability values ${\bf{V}}_{c}$  converge or the maximum number of iterations  is exhausted, then stop and proceed for decision making in following step . Otherwise, set $l=l+1$ and continue. 			
 			$$ \text{Decision:} \;\;{\hat{\bf{x}}}_c=\arg \max_{{\bf{x}}_{cm} \in {\mathbb{A}}_c} V_{c}\left({\bf{x}}_{cm}\right), \; c=1,\ldots, {JMN}/{K} $$

 	}
 
 \end{algorithm}
 
 \begin{exm}
 	Consider a $J=6$ and $K=4$  SCMA  system as shown in Fig.~\ref{fg}  and  a delay-Doppler plane  with $M=4$  and $N=4$. Each user allocates 4 SCMA codewords over one frame.   The numbers of observation nodes and variable nodes are 16 and 24 respectively.  Suppose the  observation node  $y_d$  is connected to the variable nodes $\left\{{{c}}, {{c}}_1, {{c}}_2\right\}$  through the coefficient  vectors $\left\{{\bf{h}}_{dc}, {\bf{h}}_{dc_1}, {\bf{h}}_{dc_2}\right\}$ respectively. The situation is shown in Fig.~\ref{CN_update}. Here the lengths of each coefficient vector and  symbol vector are equal to $d_v=2$.
 	\begin{figure}[!htpb] 	
 		\centering
 		\includegraphics[height=4cm, width=5.7cm]{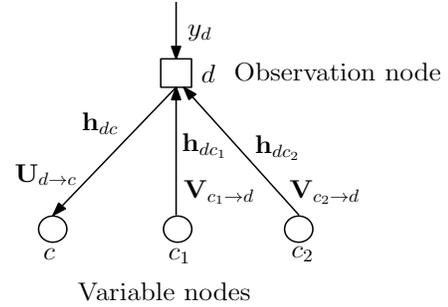}
 		\caption{Observation node update in the single-stage  MPA detection of OTFS-SCMA in uplink. }
 		\label{CN_update}
 	\end{figure}
 	 
 	 The message ${\bf{U}}_{d\rightarrow c}$ from the observation node $d$ to the user node $c$  is given by 
 \begin{equation}
 \begin{split}
 	  U_{d \rightarrow c}^{(l)}(m)  = & \sum_{\left({\bf{v}}_{c_1}, {\bf{v}}_{c_2} \right) \in {\mathbb{A}}_{c_1}\times {\mathbb{A}}_{c_2} } \frac{1}{\pi  N_0} \exp [-\frac{1}{N_0}| y_d -
 	 {\bf{h}}_{dc}{\bf{x}}_{cm} \\ 
 	 &-  {\bf{h}}_{dc_1}{\bf{v}}_{c_1}- {\bf{h}}_{dc_2}{\bf{v}}_{c_2} |^2]  V_{c_1\rightarrow d}^{(l-1)}\left({\bf{v}}_{c_1}\right)  V_{c_2\rightarrow d}^{(l-1)}\left({\bf{v}}_{c_2}\right)
 	 \end{split}
 	 \label{cnp_ex}
 \end{equation}
 Suppose, the size of each alphabet is ${{A}}=4$. Each alphabet contains $4$ codewords of length $d_v=2$.  The Cartesian product  ${\mathbb{A}}_{c_1}\times {\mathbb{A}}_{c_2} $ contains $4^2=16$ combinations of $\left({\bf{v}}_{c_1}, {\bf{v}}_{c_2} \right)$ and the outer summation in (\ref{cnp_ex}) is done over these 16 combinations. Observe that the product ${\bf{h}}_{dc_i}{\bf{v}}_{c_i}$ is a scalar complex number  and thus the argument in the exponential term in (\ref{cnp_ex}) is scalar.  
\end{exm}

The computation of the messages from the variable nodes are relatively simple. The outgoing message from a variable node to an observation node is given  by the component-wise product of the incoming messages from the other  connected observation nodes. The \textit{a posteriori} probability values of the variable nodes are computed by component-wise multiplication of all incoming messages from the neighboring observation nodes. The estimate of a  variable node is considered as that symbol in the alphabet for which the \textit{a posteriori} probability value becomes the maximum.   The details are given in Algorithm~\ref{algo:uplink}.

We present a theorem without proof regarding the  degrees of the  observation and the   variable nodes in the effective factor graph in uplink. This theorem is crucial for the complexity analysis of detection process.  
\begin{theorem}
	Consider an OTFS-SCMA model with $J\times K$ SCMA system in uplink.  Let the degrees of a variable node and a resource node in the basic SCMA system are $d_v$ and $d_f$ respectively. Let $P$ denote  the number of  multipaths  for the underlying wireless channels for each user.  Suppose for every user, the values $l_{\tau_i}, k_{\nu_i}$, $i=1,\ldots, P$ are generated randomly from $\left\{0,\ldots, P-1\right\}$.   Then for the effective factor graph of the OTFS-SCMA system free of fractional Doppler, the degrees $\left\{d_{v_i}^{\text{eff}}\right\}_{i=1}^{\frac{JMN}{K}}$ of the  variable  nodes  and the degrees $\left\{d_{f_i}^{\text{eff}}\right\}_{i=1}^{MN}$  of the observation nodes satisfy the following:
	\begin{equation}
	\begin{aligned}
	\frac{1}{\left(\frac{JMN}{K}\right)}\sum_{i=1}^{\frac{JMN}{K}}d_{v_i}^{\text{eff}}&  \leq Pd_v\\
		\frac{1}{MN}\sum_{i=1}^{MN}d_{f_i}^{\text{eff}}&  \leq Pd_f
			\end{aligned}
	\label{theorem1}
	\end{equation}
		
\end{theorem}
%
\subsection*{Complexity Analysis}
Here, the complexity of the proposed method is analyzed for downlink and uplink.  As the complexity is mainly contributed by the sophisticated detectors in the receiver side, we focus on the complexity of the detectors only. 
\subsubsection*{Downlink}
The detector in downlink is a concatenated system of LMMSE detector and MPA detector. The interference over the delay-Doppler plane is removed by the LMMSE-based OTFS detector. The LMMSE detector involves traditional matrix inversion. Its  complexity  is   ${\textit{O}}\left(M^3N^3\right)$. If the low-complexity version of LMMSE detector in \cite{Tiwari2019} is used, then the complexity would be $O\left(\frac{MN}{2}\log_2N+2MNP^2\right)$.   The output of the OTFS detector now contains the multi-user interference. The MPA-based SCMA  detector removes multi-user interference. The complexity of the SCMA detector is ${\textit{O}}\left(\frac{MN}{K}A^{d_f}\right)$ where, $A$ is the alphabet size.  For relatively small values of $M$ and $N$, the complexity of the two-stage detector is approximately   ${\textit{O}}\left(A^{d_f}\right)$.

\subsubsection*{Uplink}
In uplink, the detector comprises of only a single MPA. As per Theorem~1, the average degree  of an observation node in the effective factor graph is upper bounded by $Pd_f$, where $d_f$ is the degree of the factor node in the  basic SCMA system and $P$ is the number of  multipaths. Therefore, the complexity of the detector in uplink is ${\textit{O}}\left(MNA^{Pd_f}\right)$ as there are $MN$ observation nodes.  Note that the complexity of this detector is high if the  number of multipaths is high. If $P$ is known to be high in a particular situation, then in place of the MPA described in Algorithm~\ref{algo:uplink}, other variants based on expectation propagation \cite{EP_SCMA}, Gaussian-approximated MPA \cite{ravi_2018_TWC} etc.  may be used to reduce the complexity.

\begin{rem}
	Here we further clarify the use of two-stage and single-stage detectors in downlink and uplink respectively.  Observe from (\ref{downlink_eq_OTFS_SCMA}) that in downlink, the delay-Doppler and multi-user interactions are separable.  The matrix  ${\bf{H}}_j$  corresponds to the delay-Doppler interaction and the superimposed signal ${\bf{x}}_{\text{sum,vec}}$  arises due to multi-user fusion. On the other hand, observe from (\ref{uplink_eq})  that the delay-Doppler and multi-user interactions are not separable in uplink.  Therefore, the combined approach of OTFS and SCMA detection must be adopted. Note that this combined or single stage of detection can also be considered in downlink. However, due to high complexity of the combined detector, two stage approach is followed in downlink.   
\end{rem} 
\section{Simulation Results} \label{sec::simulations}
In this section, we present the bit-error rate (BER) simulation results for uncoded OTFS-SCMA systems. The results for both downlink and uplink are presented. In both cases, we consider OTFS with rectangular pulse. The OTFS system is assumed to free of fractional Doppler and fractional delay values. It is assumed that the receiver has the perfect knowledge of the channel.  We also compare the results of the proposed NOMA system with those of conventional OMA schemes over QAM alphabets. In order to compute the BER at any SNR, Monte Carlo simulation is carried out for $5\times 10^{4}$ frames. 
\subsection{Downlink}{\label{sec::downlink}}
First we consider a delay-Doppler plane $\Gamma_{8,8}$.  An SCMA system with $J=6$ and $K=4$ is considered.  The SCMA codebooks proposed in \cite{zhang2016capacity} are used. The overloading factor of the OTFS-SCMA system is $\lambda=J/K=150\%$. 
The BS transmits $MN/K=16$ symbols for each user. First, as per Scheme~1, 16 SCMA codewords of every user are allocated on respective delay-Doppler plane.  
The BS then superimposes the input signals over these 6 delay-Doppler grids. The superimposed signal is fed to the  OTFS modulator which comprises of ISFFT and Heisenberg transforms.  The modulator output   is transmitted over the wireless channel. 

\begin{figure}[!ht] 	
		\centering
		\includegraphics[height=9cm, width=9cm]{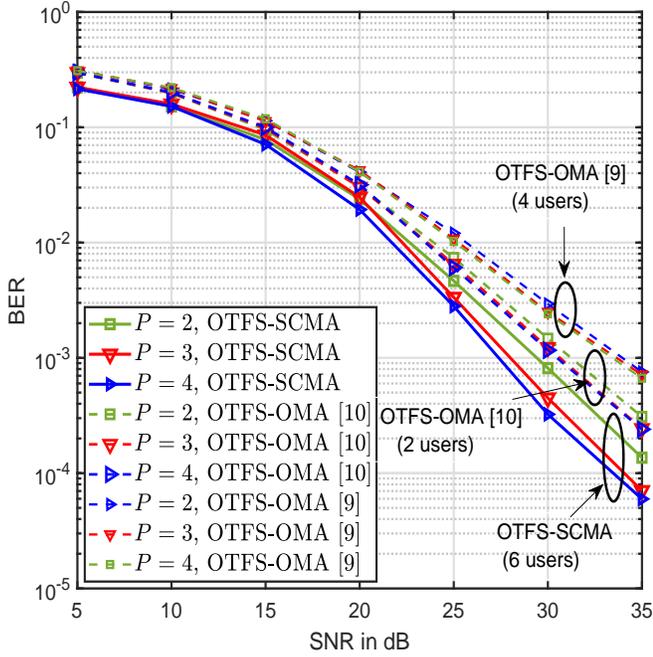}

		\caption{OTFS-SCMA with $N=8$, $M=8$ and ($J=6$, $K=4$) SCMA system, overloading factor 150$\%$, (Scheme~1) in downlink.}
		\label{sim_down1}
	\end{figure}
%
	The wireless channel is represented by  $P$ propagation paths.   The $i^{\text{th}}$  path is associated with integer delay ($l_{\tau_i}$) and Doppler (${k}_{\nu_i}$)  components as explained in Section~\ref{sec::TF_des}. We consider $\triangle f=1$ and  $T=1$.  One of $P$ paths is the line-of-sight (LoS) path with ($l_{\tau_i}=0,{k}_{\nu_i}=0 $). The  remaining values of  $l_{\tau_i}$s and  ${k}_{\nu_i}$s are   generated randomly from $\left\{0,\ldots, P-1\right\}$. In our simulations, we have considered distinct propagation paths with distinct pairs of ($l_{\tau_i},{k}_{\nu_i} $). 
	
	The received signal is passed through the OTFS demodulator, OTFS LMMSE detector and SCMA detector block sequentially as shown in Fig.~{\ref{block_downlink}}. The BER performance of this OTFS-NOMA scheme is presented in Fig.~\ref{sim_down1}.  The results for $P=2, 3,4$ are shown. Observe that as $P$ increases,  the   BER curve improves. This improvement may be attributed to the increased diversity in the system \cite{ravi_diversity,surabhi_diversity}.   
	\begin{figure}[!ht]			
		\centering		
		\includegraphics[height=5cm, width=8cm]{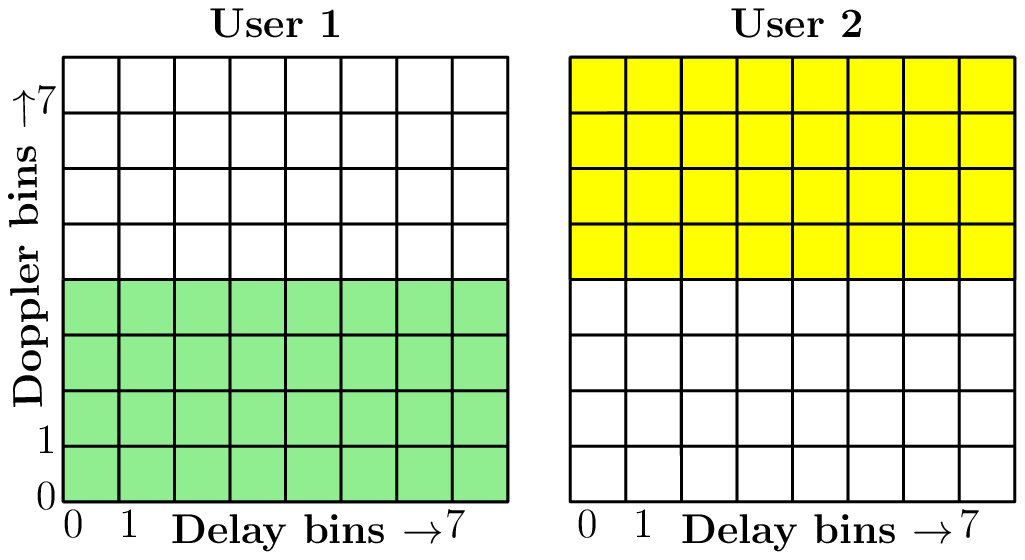}
		\caption{OTFS-OMA with $N=8$, $M=8$ with two users \cite{surabhi_MA} in downlink.}
		\label{OMA_DD1} 
	\end{figure}
	
\begin{figure}[!ht]			
		\centering		
		\includegraphics[height=3.8cm, width=6cm]{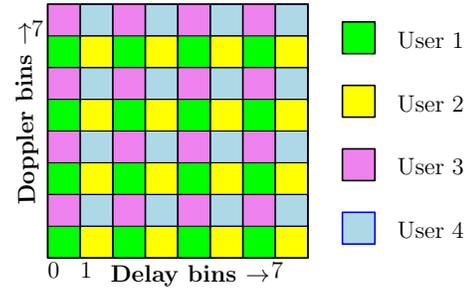}
		\caption{OTFS-OMA with $N=8$, $M=8$ with four users \cite{Khammammetti2019a} in downlink.}
		\label{OMA_DD2} 
	\end{figure}

	Fig.~\ref{sim_down1} also presents results for OTFS-OMA over $\Gamma_{8,8}$. The  OTFS-OMA scheme proposed in \cite{surabhi_MA} is considered first.  We   consider two users occupying the delay-Doppler plane as shown in Fig.~\ref{OMA_DD1}. Each user's QAM symbols are placed in the plane in a non-overlapping manner as depicted in Fig.~\ref{OMA_DD1}.   Observe that the number of symbols that can be transmitted by each user is $32$.  Thus  for two users, the BS transmits 64 symbols over 64 slots in the plane. The overloading factor is 100$\%$. LMMSE-based detector is used after OTFS demodulator.  We have also considered the OTFS-OMA method proposed in \cite{Khammammetti2019a} for 4 users. The symbols of the 4 users are placed on $\Gamma_{8,8}$ as shown in Fig.~\ref{OMA_DD2}. Here also, the overloading factor is 100$\%$. 
	
	 The BER results for these OMA schemes with $P=2,3,4$ are shown in Fig.~\ref{sim_down1}. We consider QAM alphabet as its size $(A=4)$ is exactly same as that of the  OTFS-SCMA system under consideration.  Observe that, even though the OTFS-SCMA system involves 6 users with overloading factor $150\%$,  it performs better than the OTFS-OMA schemes with  2 users and 4 users with  overloading factor of 100$\%$. Remark~\ref{noma_remark}  
	 explains the performance improvement of OTFS-SCMA over OTFS-OMA. 
 	
 	Next, we consider an SCMA system with $J=8$  and $K=4$.  The overloading factor of this OTFS-SCMA scheme is $200\%$. The $4\times 6$ factor matrix $\bf{F}$ shown in (\ref{eq:F_matirx}) is enlarged to the following $4 \times 8$  factor matrix ${\bf{F}}_e$:
 	 	\begin{equation}
 	 	\label{expanded_F}
 			{\bf{F}}_e=
 			\begin{bmatrix}
 			1 & 0 & 1 & 0 & 1 & 0 & 1& 0\\
 			0 & 1 & 1 & 0 & 0 & 1 & 1& 0\\
 			1 & 0 & 0 & 1 & 0 & 1 & 0& 1\\
 			0 & 1 & 0 & 1 & 1 & 0 & 0& 1\\
 			\end{bmatrix}	
 	\end{equation}
 	Note from (\ref{expanded_F}) that ${\bf{F}}_e$ is obtained from ${\bf{F}}$ by appending the $3^{\text{rd}}$ and the $4^{\text{th}}$ columns of ${\bf{F}}$ to itself sequentially. In ${\bf{F}}_e$, the $7^{\text{th}}$  and $8^{\text{th}}$ columns are identical to the $3^{\text{rd}}$ and the $4^{\text{th}}$ respectively. Observe from (\ref{expanded_F}) that the $3^{\text{rd}}$ and the $4^{\text{th}}$ columns are non-overlapping.  Therefore, the  codebooks for the $7^{\text{th}}$ and  the $8^{\text{th}}$ users  are  set exactly to  those for the $4^{\text{th}}$ and the $3^{\text{rd}}$ users respectively. 	These codebooks are used for the OTFS-SCMA modulation. 
		\begin{figure}[!ht] 	
			\centering
			\includegraphics[height=7.5cm, width=8.3cm]{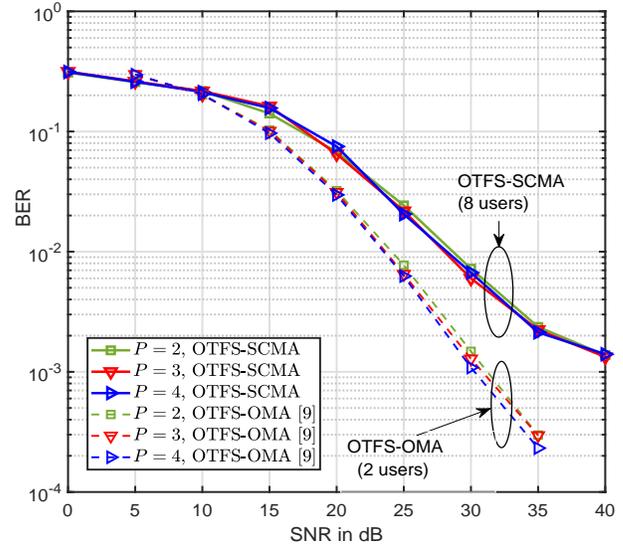}

			\caption{OTFS-SCMA with $N=8$, $M=8$ and ($J=8$, $K=4$) SCMA system, overloading factor 200$\%$ in downlink.}
			\label{sim_down2}
		\end{figure}
	Fig.~\ref{sim_down2} presents the BER performance of the OTFS-SCMA system with Scheme 1 in downlink. Although the overloading factor is 200$\%$, the BER performance is impressive. This performance is compared with OTFS-OMA method \cite{Khammammetti2019a} for 2 users. Note that the performance of the OTFS-SCMA with an overloading factor of 200$\%$  is worse than OTFS-OMA with 2 users. The plots in Fig.~\ref{sim_down2} shows that the BER performance is almost similar for $P=2,3,4$. This observation may arise due to excessive overloading and repetition of a few  codebooks amongst users. 
	
\begin{figure}[!ht] 	
		\centering
		\includegraphics[height=9cm, width=8.3cm]{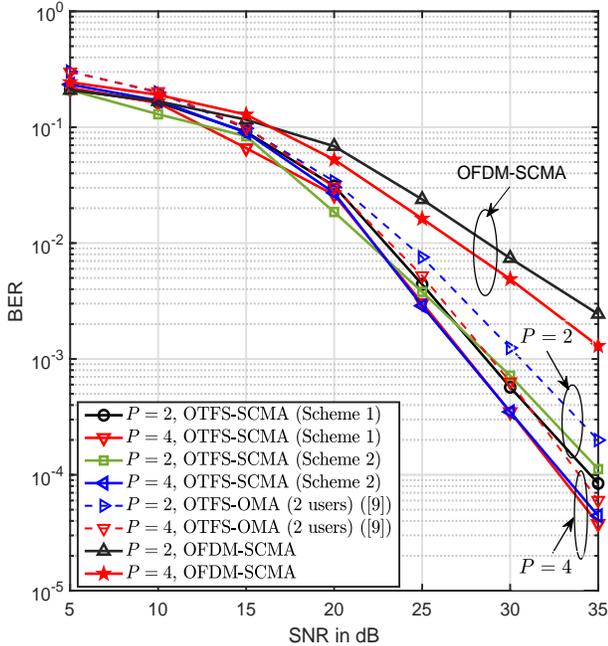}

		\caption{OTFS-SCMA with $N=8$, $M=16$ and ($J=6$, $K=4$) SCMA system (Scheme 1 and Scheme 2) in downlink.}
		\label{sim_down_schemes_P2P4}
	\end{figure}

	Now we consider a delay-Doppler plane $\Gamma_{M,N}$ with unequal values of $M$ and $N$  and compare the performances of Scheme~1 and Scheme~2 which are described in Section~\ref{sec::downlink}.  We consider $M=16$  and $N=8$. A $6\times4$  SCMA system  is adopted.  Fig.~\ref{sim_down_schemes_P2P4} presents the BER performances of the schemes for $P=2$ and $P=4$.  For $P=2$, Scheme~1 performs slightly better than Scheme~2. However, as $P$ increases, it has been observed that the performances of both Scheme 1 and Scheme 2 are similar.  
	
	Fig.~\ref{sim_down_schemes_P2P4} also shows the performance of the OFDM-SCMA system.  The same simulation set-up of OTFS-SCMA  is considered here. The delay-Doppler channel representation is used.   The only difference is that the information symbols are directly put on the time-frequency  domain. In other words, the ISFFT and the SFFT transformations are not considered in the modulator and the demodulator respectively.  Observe that the performance of OFDM-SCMA is not satisfactory. There is a significant performance gain of  OTFS-SCMA over OFDM-SCMA.  This observation in the multi-user scenario  is in agreement with the result reported in \cite{ravi_2018_TWC} for the single-user case. 
\subsection{Uplink}
Here we present the simulation results for OTFS-SCMA  with different values of $M$ and $N$ in uplink.  First we  consider  delay-Doppler plane with $M=4$, $N=4$ and $J=6, K=4$ SCMA system.  In this section, Scheme~1 is used for allocating the SCMA codewords over the delay-Doppler plane. 
	\begin{figure}[!ht] 	
		\centering
		\includegraphics[height=7.5cm, width=8.3cm]{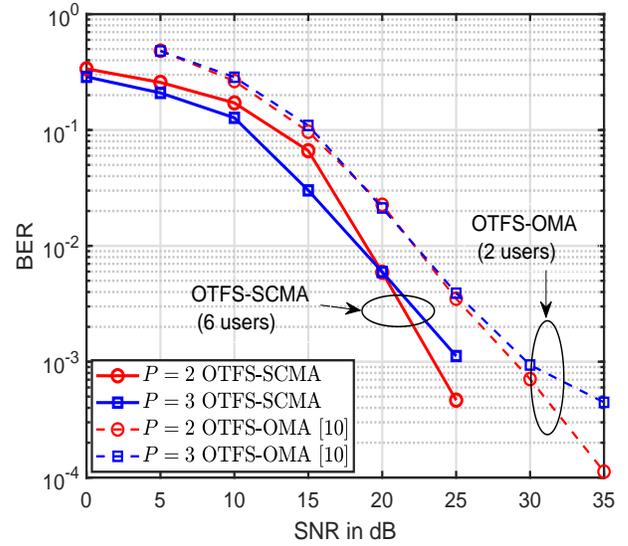}

		\caption{OTFS-SCMA with $N=4$, $M=4$ and ($J=6$, $K=4$) SCMA system  and OTFS-OMA over QAM  in uplink. }
		\label{sim_up1}
	\end{figure}
	Each of the  6 users places 4 SCMA codwords  over the delay-Doppler plane $\Gamma_{4,4}$. Thus the overloading factor becomes $\lambda=24/16=150\%$. The respective output of the OTFS modulator is transmitted over  the channel towards the BS.    The BS receiver first feed the signal from the channel to an OTFS demodulator. The effective coefficient matrix  $\bf{H}$ is found out as explained in Section~\ref{sec::uplink_prop}.  Then using this matrix, the single-stage MPA-based detection is carried out to extract  the individual users' data symbols. Note that this detector acts as the combined detector of OTFS and SCMA symbols. 
	Fig.~{\ref{sim_up1} shows the BER plots for the proposed OTFS-SCMA scheme for $P=2$  and $P=3$.   The results of OTFS-OMA \cite{surabhi_MA} with 2 users are also shown in Fig.~\ref{sim_up1}. 4-QAM is considered for OTFS-OMA.  The QAM  symbols are placed in the delay-Doppler grid $\Gamma_{4,4}$ in the same way as depicted in Fig.~\ref{OMA_DD1}. For  OTFS-OMA in uplink, the detection is carried out with the help of MPA.   Note that  in uplink too, the OTFS-SCMA performs significantly better than its OMA counterpart.  Moreover, the performance of the proposed  OTFS-SCMA is  better in uplink than in downlink.  This performance difference may be accredited to the combined powerful MPA detector in uplink compared to the two stage detection process in downlink.   Note that the BER  for $P=3$ tends to be worse than that for $P=2$ after a certain SNR value for both OTFS-SCMA and OTFS-OMA.  This peculiar transition happens as   the value of $P=3$ is close to the number of delay $(M)$ and Doppler ($N$) bins. 
	
	\begin{figure}[!ht] 	
		\centering
		\includegraphics[height=7.5cm, width=8.3cm]{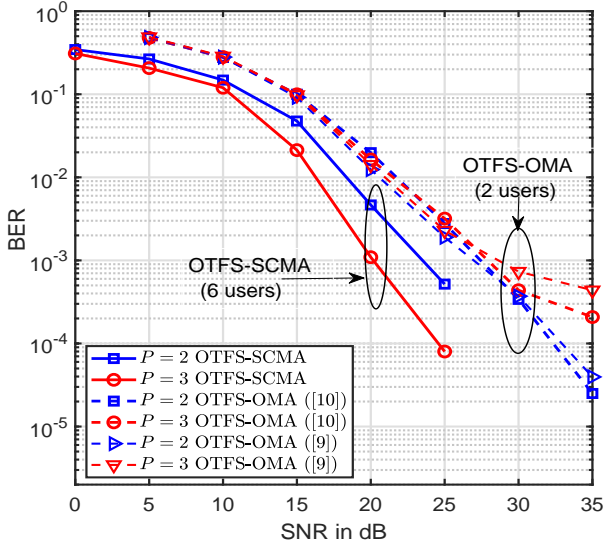}

		\caption{OTFS-SCMA with $N=4$, $M=8$ and ($J=6$, $K=4$) SCMA system  and OTFS-OMA over QAM  in uplink.}
		\label{sim_up2}
	\end{figure}
	
	Next we consider a delay-Doppler grid $\Gamma_{4,8}$  with $J=6, K=4$  SCMA system. Fig.~\ref{sim_up2} presents the BER plots for the proposed OTFS-SCMA system for $P=2$  and $P=4$. Observe that as $P$ increases the BER improves due to the enhancement of the diversity.   The results of OTFS-OMA \cite{Khammammetti2019a,surabhi_MA} are also shown in Fig.~\ref{sim_up2}. Here also, the OTFS-SCMA performs better than OTFS-OMA over the alphabets of same size. 
	
	\begin{rem} \label{noma_remark}
		Observe from Fig.~\ref{sim_down1}, Fig.~\ref{sim_down_schemes_P2P4},  Fig.~\ref{sim_up1} and Fig.~\ref{sim_up2} that the proposed OTFS-SCMA method outperforms the conventional OTFS-OMA schemes \cite{surabhi_MA,Khammammetti2019a} in both downlink and uplink.
			Note that the alphabets of OTFS-SCMA system  contain codeword vectors of length $K=4$ for a $6\times 4$ SCMA system.  The detection is done vector-wise. Moreover, SCMA codebooks can offer impressive coding and shaping gain \cite{nikopour2013}.  On the other hand, in OTFS-OMA, the alphabet is the standard 4-QAM where any codeword is a scalar complex. These facts contribute towards the   performance gain of OTFS-SCMA over OTFS-OMA. 
	\end{rem}
\section{Conclusions}  \label{sec::conc}
This paper presented a novel OTFS-NOMA  strategy based on SCMA.  This proposed method considers  pre-designed SCMA codebooks with a particular overloading factor.   The sparse SCMA codewords of each  user are placed on the delay-Doppler grid either vertically or horizontally.  The ratio of the  total number of codewords or symbols transmitted by all users to the total number slots or resources in the  delay-Doppler plane   is exactly equal to the overloading factor of the underlying basic SCMA system. The proposed  OTFS-SCMA scheme has been devised and analyzed for both downlink and uplink.  The receiver in downlink includes a two stage detection process.  First,   the intermingling of the symbols in different delay-Doppler slots is resolved with the help of an OTFS detector.  Subsequently, an SCMA detector eliminates   the multi-user interaction.     In uplink, the receiver in the BS comprises of a combined OTFS-SCMA detector based on a single-stage of MPA.  Both delay-Doppler and multi-user interactions are resolved at one shot.  Simulation results are presented with different parameters for both the cases of downlink and uplink with overloading factor upto 200$\%$. 
The results showed that the proposed OTFS-SCMA system can yield better BER performances than the conventional OTFS-OMA systems.  


\vspace{-0.02in}
\bibliographystyle{ieeetran}
\footnotesize
\bibliography{references_OTFS_SCMA}

\end{document}